\documentclass[twocolumn,english,reprint,superscriptaddress,prl]{revtex4-1}

\usepackage[T1]{fontenc}
\usepackage[latin9]{inputenc}
\usepackage{color}
\usepackage{bm}
\usepackage{amstext}
\usepackage{amssymb}
\usepackage{graphicx}
\usepackage{booktabs}
\usepackage{tabularx}
\usepackage{amsmath}
\makeatletter

\begin{document}
\title{Corner entanglement of a resonating valence bond wavefunction}

\author{Giacomo Torlai}
\affiliation{AWS Center for Quantum Computing, Pasadena, CA, USA}

\author{Roger G. Melko}
\affiliation{Department of Physics and Astronomy, University of Waterloo, Ontario, N2L 3G1, Canada}
\affiliation{Perimeter Institute for Theoretical Physics, Waterloo, Ontario, N2L 2Y5, Canada}
\date{\today}

\begin{abstract}
We perform a quantum Monte Carlo simulation of the resonating valence bond wavefunction on a two-dimensional square lattice with periodic boundary conditions. 
Using two replicas of the system, we calculate the second Renyi entropy on a spatial bipartition with a square geometry.
Through a finite-size scaling analysis, we extract the logarithmic correction to the area law due to the presence of the sharp corners in the entangling surface. We find that the coefficient of this logarithm is positive with a value of $0.073$ for a single $90^{\circ}$ corner.
\end{abstract}

\maketitle

{\it Introduction.} The second Renyi entropy $S_2$ provides a powerful window on the behavior of entanglement in interacting systems, since it can be calculated with statistically exact quantum Monte Carlo (QMC) methods on finite size systems \cite{PhysRevLett.104.157201}.  
One of the most important jobs for large-scale QMC simulations is to study scaling properties of the entanglement at quantum critical points.
In particular, universal quantities in the entanglement entropy that emerge as subleading corrections to the area law can require very large lattice sizes to resolve accurately \cite{PhysRevB.86.235116,Inglis_2013}.
In this paper, we perform QMC simulations on the two-dimensional ($2D$) resonating valence bond (RVB) wavefunction \cite{PhysRevB.82.180408,PhysRevB.84.174427}, and extract $S_2$
for a square bipartition for lattice sizes with up to $68 \times 68$ sites.
We observe a clear logarithmic correction to the area law due to the four vertices of the square bipartition. Performing a careful finite-size scaling analysis of our data, we find that the coefficient of this logarithm is positive with a value of $0.073$ for each $90^{\circ}$ corner.

{\it RVB wavefunction.}
We define the resonating valence bond (RVB) wavefunction on a square lattice with a total number of sites $N=L^2$ and periodic boundary conditions
\begin{equation}
|\psi\rangle = \sum_\alpha |\phi_\alpha\rangle\:,
\end{equation}
where $|\phi_\alpha\rangle$ is a nearest-neighbor valence bond (VB) state\begin{equation}
|\phi_\alpha\rangle=2^{-N/4}\sum_{j=1}^{N/2}\big(|\uparrow_j\:\downarrow_{\alpha_j}\rangle-|\downarrow_j\:\uparrow_{\alpha_j}\rangle\big)\:.
\end{equation}
Here the index $j$ and $\alpha_j$ run over sites on different sublattices. 

\begin{figure}[t]
\noindent \centering{}\includegraphics[width=0.8\columnwidth]{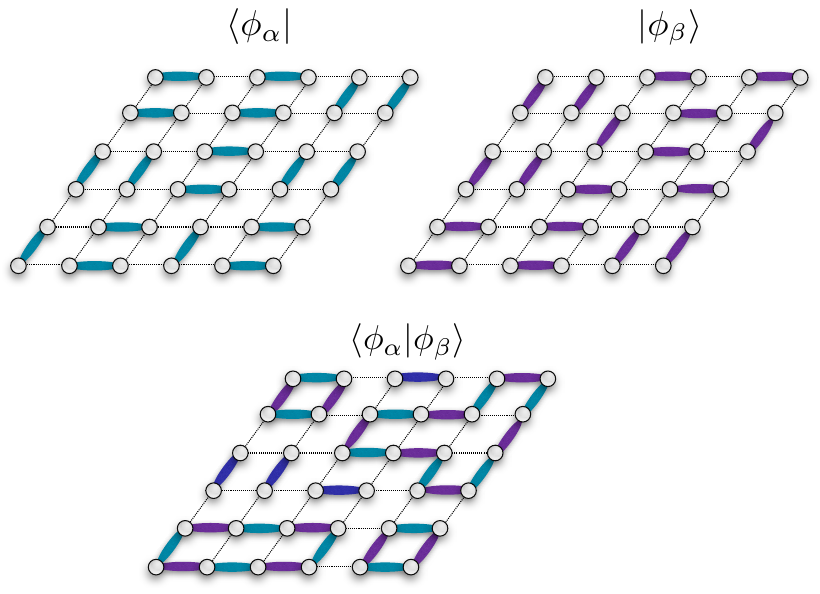}
\caption{The SU(2) RVB wavefunction. At top are two different VB states, with their transition graph at bottom. The number of closed loop is equal to $n_{\alpha\beta}=8$, resulting into an overlap of $\langle\phi_\beta|\phi_{\alpha}\rangle=2^{-10}$.}
\label{Fig::rvb} 
\end{figure}

The set of all $(N/2)!$ possible VB states forms an overcomplete and non-orthonormal basis of the Hilbert space. In order to calculate the overlap between two VB states, $\langle\phi_\alpha|\phi_{\beta}\rangle$, one first re-writes the states in the spin basis using the Marshall-sign rule, e.g.
\begin{equation}
|\phi_\alpha\rangle=2^{-N/4}\sum_{j=1}^{2^{N/2}}(-1)^{N_\alpha^\uparrow}|\bm{S}_{\alpha j}^z\rangle
\end{equation}
where $N_\alpha^\uparrow$ is the number of up-spins in one sublattice of reference. The overlap can be then evaluated by considering the transition graph between the two VB states (Fig. \ref{Fig::rvb}). Following the orthonormality condition of the spin basis, there are two possible spin states for each closed loop in the transition graph, corresponding to a chain of anti-parallel spins. Therefore, the calculation of the overlap translates into counting the number $n_{\alpha\beta}$ of loops in the graph,
\begin{equation}
\langle\phi_\alpha|\phi_{\beta}\rangle = 2^{n_{\alpha\beta}-N/2}\:.
\end{equation}

We now turn to the calculation of the expectation value of a generic observable $\mathcal{O}$ in the RVB wavefunction $|\psi\rangle$. This is simply given by
\begin{equation}
\begin{split}
\langle\mathcal{O}\rangle&=
\frac{\sum_{\alpha\beta}\langle\phi_\beta|\mathcal{O}|\phi_\alpha\rangle}{\sum_{\alpha\beta}\langle\phi_\beta|\phi_\alpha\rangle}=
\frac{\sum_{\alpha\beta}\langle\phi_\beta|\phi_\alpha\rangle\frac{\langle\phi_\beta|\mathcal{O}|\phi_\alpha\rangle}{\langle\phi_\beta|\phi_\alpha\rangle}}{\sum_{\alpha\beta}\langle\phi_\beta|\phi_\alpha\rangle}\\
&\equiv\frac{1}{\sum_{\alpha\beta}W_{\alpha\beta}}\sum_{\alpha\beta}W_{\alpha\beta}\frac{\langle\phi_\beta|\mathcal{O}|\phi_\alpha\rangle}{\langle\phi_\beta|\phi_\alpha\rangle}\\
&=\bigg\langle\frac{\langle\phi_\beta|\mathcal{O}|\phi_\alpha\rangle}{\langle\phi_\beta|\phi_\alpha\rangle}\bigg\rangle_{W_{\alpha\beta}}\:.
\end{split}
\end{equation}
Since $\langle\phi_\beta|\phi_\alpha\rangle>0\,\,\forall\alpha,\beta$, we can regard $W_{\alpha\beta}$ as statistical weights and approximate the expectation value  $\langle\mathcal{O}\rangle$ as a Monte Carlo (MC) estimator,
\begin{equation}
\langle\mathcal{O}\rangle\simeq\frac{1}{M}\sum_{k=1}^M\frac{\langle\phi_{\beta_k}|\mathcal{O}|\phi_{\alpha_k}\rangle}{\langle\phi_{\beta_k}|\phi_{\alpha_k}\rangle}\:,
\end{equation}
where the states $\phi_{\alpha_k}$ and $\phi_{\beta_k}$ are drawn from the distribution $W_{\alpha\beta}$. Given some update procedure $(\alpha,\beta)\rightarrow(\alpha^\prime,\beta^\prime)$ satisfying detailed balance, the acceptance rate is given as usual by the ratio $W_{\alpha\beta}/W_{\alpha^\prime\beta^\prime}=2^{n_{\alpha\beta}-n_{\alpha^\prime\beta^\prime}}$, which involves loop counting in the transition graph. In order to avoid this step, which can become computationally cumbersome in the presence of long-range Neel order (i.e. very long loops), an efficient sampling algorithm was proposed in Ref.~\cite{PhysRevB.84.174427} which we briefly summarize here. 

\begin{figure}[t]
\noindent \centering{}\includegraphics[width=0.7\columnwidth]{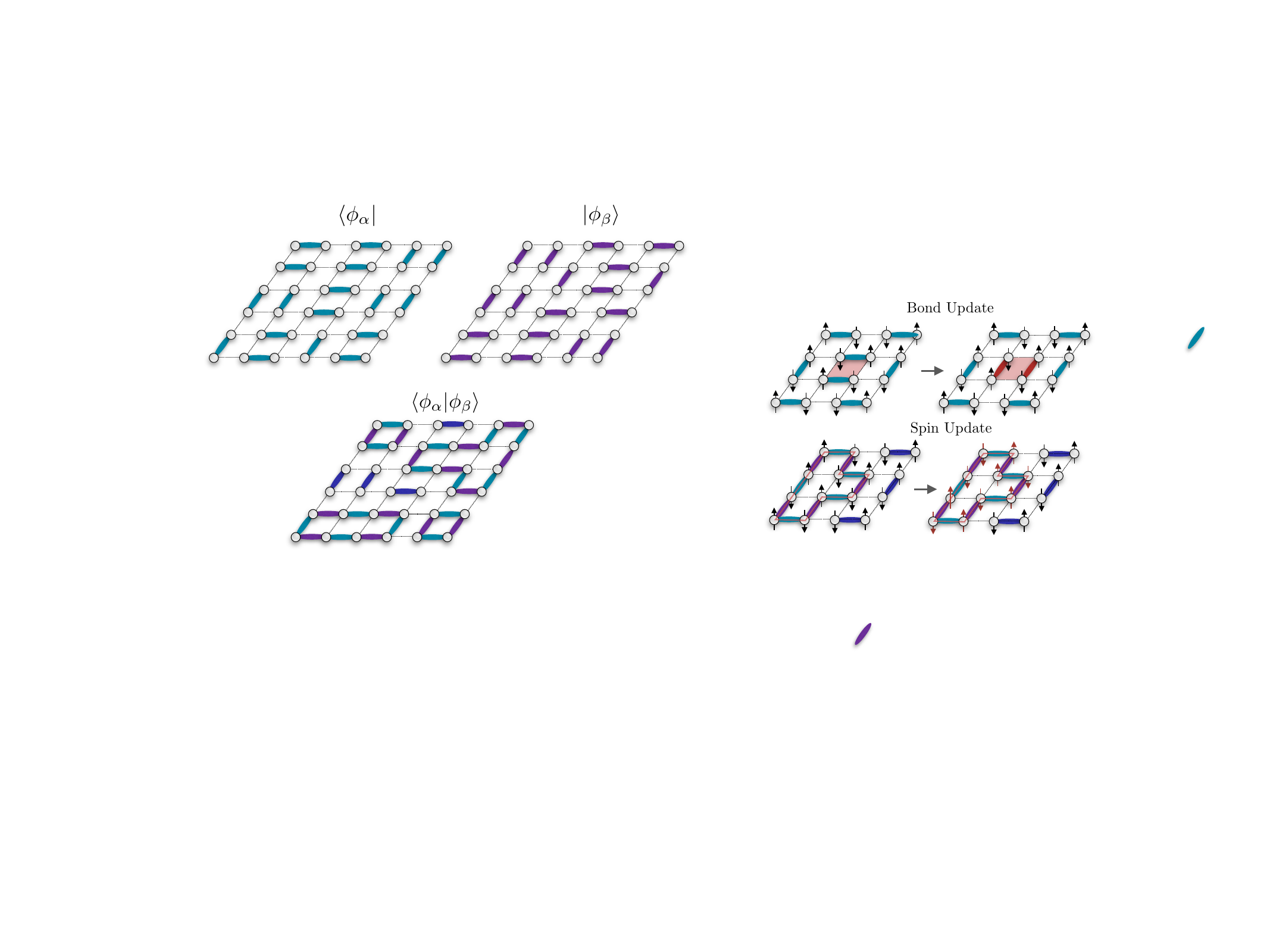}
\caption{Monte Carlo updates. In the bond update (top), two VB states within a given plaquette are exchanged, provided the spins are compatible with the new configuration. In the spin update (bottom) all the spins for a given closed loop in the transition graph are flipped with probability $1/2$.}
\label{Fig::mc} 
\end{figure}

At any given simulation step, the VB state (e.g.~the ket state) is written as $|\phi_\alpha\rangle = |B_\alpha\rangle\otimes|Z_\alpha\rangle$, where $|B_\alpha\rangle$ represents the bond configurations, i.e. the dimer coverings of the full square lattice, and $|Z_\alpha\rangle=|S_{1\alpha}^z,\dots,S_{N\alpha}^z\rangle$ is a spin product state compatible with the bond state $|B_\alpha\rangle$ (Fig. \ref{Fig::mc}a). A similar representation holds for the bra state $\langle\phi_\beta|=\langle B_\beta|\otimes\langle Z_\beta|$, but since we require that $W_{\alpha\beta}\ne 0$, it follows that the two spin state must the identical at all times, $|Z_\alpha\rangle=|Z_\beta\rangle\equiv|Z_{\alpha\beta}\rangle$. Therefore, a single MC step consists in proposing two new bond configurations $|B_{\alpha^\prime}\rangle$,$|B_{\beta^\prime}\rangle$ for the bra and the ket respectively, as well as a new spin configuration $|Z^\prime_{\alpha^\prime\beta^\prime}\rangle$ compatible with both states. The simplest bonds update exchanges bonds between two sites in the same sublattice. This update is always accepted if the new bond configuration is compatible with the spin state (Fig. \ref{Fig::mc}a-b), and rejected otherwise. 
After the bond configuration in the bra and ket are updated, a new spin configuration is sampled by building the transition graph and randomly flipping all the spins within a given loop with probability $1/2$. 


{\it Reyni entanglement entropy.}
Given a bipartition of the physical system into a region $A$ and its complement $\bar{A}$, the second Renyi entanglement entropy is defined as
\begin{equation}
S_2(A)=- \log\text{Tr} \big[\rho^2_A \big],
\end{equation}
where 
$
\rho_A = \text{Tr}_{\bar{A}}|\psi\rangle\langle\psi|
$
is the reduced density matrix for region $A$. $S_2(A)$ is directly related to the expectation value of the ``swap'' operator between two replicas of the original wavefunction,
\begin{equation}
e^{-S_2}=\text{Tr}\rho^2_A=\langle Swap_A\rangle=\frac{\langle\Psi|Swap_A|\Psi\rangle}{\langle\Psi|\Psi\rangle}\:.
\end{equation}
Here,
$
|\Psi\rangle=\sum_{\alpha_1\alpha_2}|\phi_{\alpha_1}\rangle\otimes|\phi_{\alpha_2}\rangle\:
$
is the wavefunction of the two copies of the original system and $Swap_A$ acts on $|\Psi\rangle$ by exchanging configurations of region $A$ between the two replicas (while leaving region $\bar{A}$ unchanged). The expectation value of the swap operator can be computed as the average
\begin{equation}
\langle Swap_A\rangle=\bigg\langle\frac{\langle\phi_{\beta_1}\phi_{\beta_2}|Swap_A|\phi_{\alpha_1}\phi_{\alpha_2}\rangle}{\langle\phi_{\beta_1}\phi_{\beta_2}|\phi_{\alpha_1}\phi_{\alpha_2}\rangle}\bigg\rangle_{W_{\alpha_1\alpha_2;\beta_1\beta_2}}
\end{equation}
over the probability distribution 
\begin{equation}
W_{\alpha_1\alpha_2;\beta_1\beta_2}=\langle\phi_{\beta_1}\phi_{\beta_2}|\phi_{\alpha_1}\phi_{\alpha_2}\rangle=2^{n_{\alpha_1\beta_1}+n_{\alpha_2\beta_2}-N}\:.
\end{equation}
\begin{figure}[t]
\noindent \centering{}\includegraphics[width=\columnwidth]{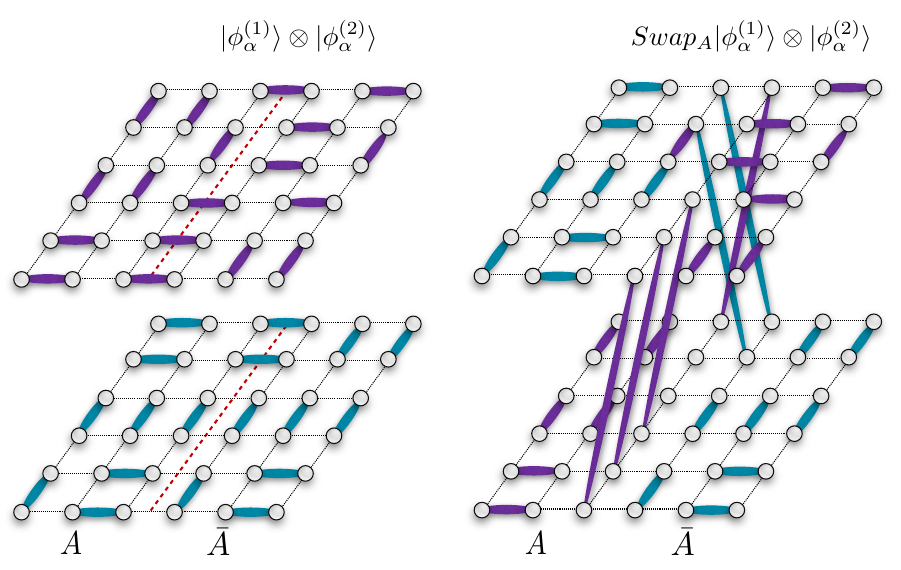}
\caption{Measurement of the swap operator. The QMC simulation is carried out over two replicas of the RVB wavefunction (left). The swap operator acts on the composite system by swapping the configuration of the degrees of freedom inside the region $A$ (right).}
\label{Fig::replica} 
\end{figure}

In practice for our simulations, four RVB wavefunctions can be sampled, i.e.~two replicas of the bra and ket state respectively. The replicas of each state can be sampled independently using the sampling scheme presented in the previous section, i.e.~local bond updates are performed on each of the four bond states $\{|B_{\alpha_1}\rangle,|B_{\alpha_2}\rangle,|B_{\beta_1}\rangle,|B_{\beta_2}\rangle\}$, followed by spin updates on the two spin states $\{|Z_{\alpha_1\beta_1}\rangle,|Z_{\alpha_2\beta_2}\rangle\}$. These configurations are then used to calculated the expectation value as the MC average
\begin{equation}
\langle Swap_A \rangle \simeq\frac{1}{M}\sum_{k=1}^N 2^{n_{\alpha\beta}^{swap}-n_{\alpha\beta}}
\end{equation}
where $n_{\alpha\beta}$ and $n_{\alpha\beta}^{swap}$ are the total number of loops in the transition graphs between the two replicas for the unswapped and swapped configurations respectively.
Finally, we use a standard ``improved ratio'' sampling to improve our statistics, as described in Ref.~\cite{PhysRevLett.104.157201}, where the region $A^{i}$ is built up from a number of independent simulations of smaller geometric regions $A^{<i}$.  In the below results, we constrain $A^{i-1}$ and $A^{i}$ to differ by at most 4 lattice sites.

%
%


\begin{figure}[t]
\noindent \centering{}\includegraphics[width=\columnwidth]{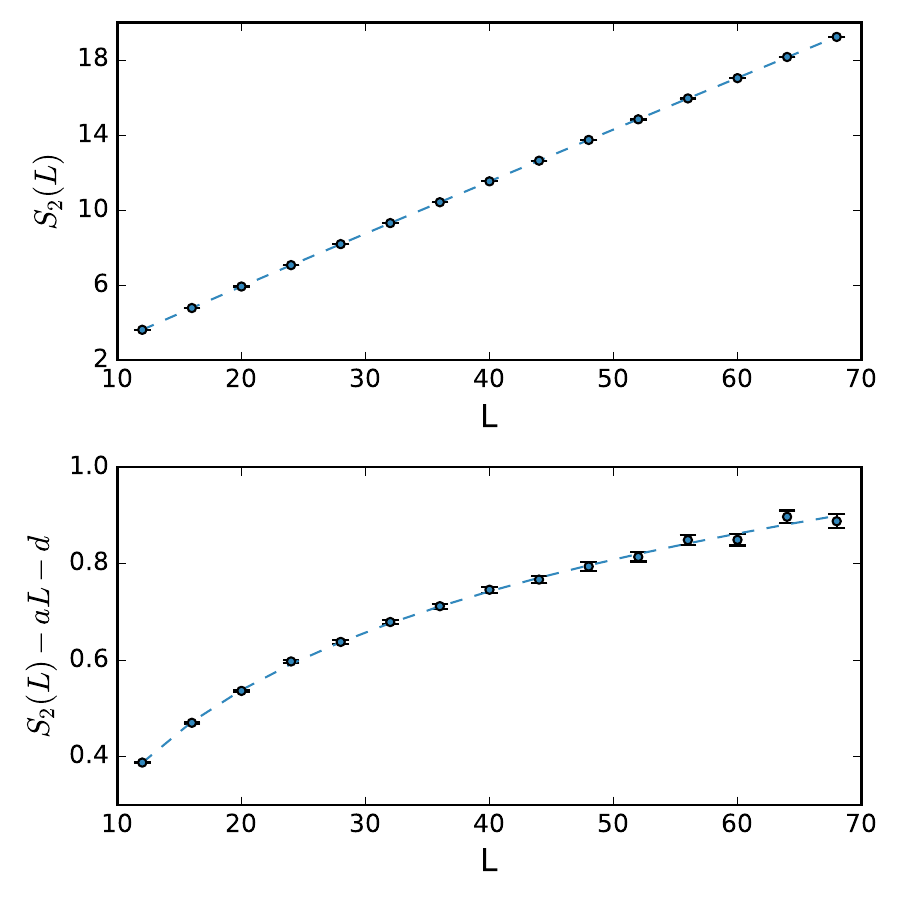}
\caption{Scaling of the second Renyi entropy with the size of the system $L$, for a fixed aspect-ratio of $L/2$ for the entangling subregion.}
\label{Fig::S2} 
\end{figure}

\begin{figure}[t]
\noindent \centering{}\includegraphics[width=\columnwidth]{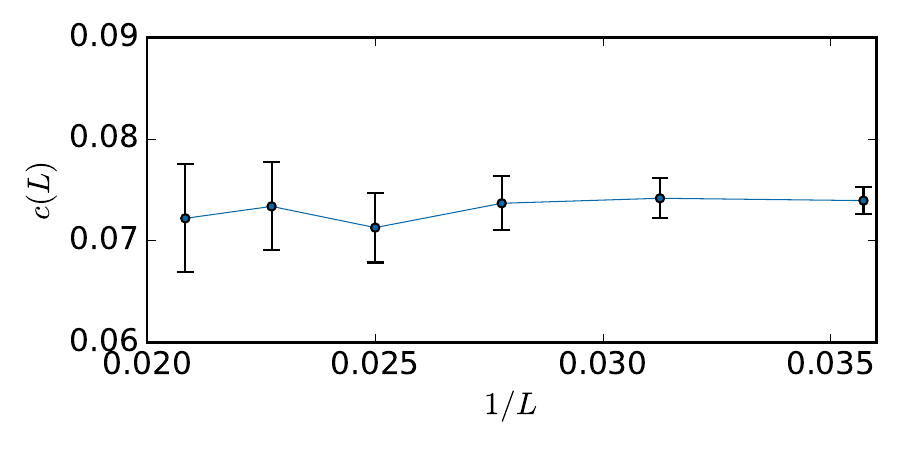}
\caption{Finite-size scaling of the corner coefficient.}
\label{Fig::corner} 
\end{figure}

{\it Results.}
The second Renyi entropy of the RVB wavefunction in $2D$ has an area law, with subleading corrections that depend on the geometry of the bipartition $A$ \cite{PhysRevB.85.165121,Stephan_2013}.
We simulate the RVB wavefunction for square periodic lattices with linear size $L\in[12,68]$, and measure the entanglement entropy for a subregion $A$ consisting of a square with linear size $L_A=L/2$. Adjacent geometries in the improved ratio expansion differ at most by $4$ sites. 

We use finite-size scaling to analyze our Renyi entropy data, assuming a function form,
\begin{equation}
S_2(L)=aL+4c\log L + d.
\label{Eq::S2_L}
\end{equation}
Here, $a$ is the non-universal area law coefficient, $4c$ the contribution from the four corners, and $d$ is a non-universal constant.
Using our entire data range, we plot the scaling of $S_2(L)$ in Fig.~\ref{Fig::S2} (top), showing the dominance of the area law contribution. A fit using the functional form from Eq.~\eqref{Eq::S2_L} generates a universal coefficient 
$c=0.073(2)$ with $\chi^2=0.958$. In the bottom panel of Fig.~\ref{Fig::S2} we isolate a clear logarithmic signal from the four corners, by subtracting the scaling parameters $aL$ and $d$ found in the previous fit.

Finally, we examine the dependence of the corner coefficient on the system sizes included in the fit. We characterize the size dependent $c(L)$ using a sliding window over system sizes $L$. Namely, we select entanglement entropy data from a subset $L\in[L_{min},L_{max}]$, containing 10 consecutive system sizes $L$. Starting from the lowest $L$ in our simulations, we run the above fitting routine for consecutive windows, leading to a total of 6 data points for $c(L)$.
We show the result in Fig.~\ref{Fig::corner}, where $L$ is defined as the midpoint of each window. As is clear from this figure, we observe no significant dependence of $c$ on $L$ using this procedure.

{\it Discussion.} In this paper, we have shown that the RVB wavefunction in $2D$ has a logarithmic correction to the area law due to the presence of the sharp corners in the entangling surface. 
Using finite-size QMC simulations to access the second Renyi entropy, we determine that the coefficient of this logarithm is positive with a value of $0.073$ for a single $90^{\circ}$ corner.
This positive value stands in contrast the sign of the logarithmic correction induced by sharp vertices in similar entangling surfaces in conformal field theories \cite{TomoyoshiHirata_2007}. We hope our results will help inform theoretical studies of other critical systems with positive corner coefficients, particularly in models with candidates for deconfined quantum critical points \cite{PhysRevLett.128.010601,plusA2,plusA3,plusA4}.

{\it Acknowledgements.} We are indebted to a large number of friends for many conversations about this calculation over the last few years, including
M. Cheng,
P. Fendley,
T. Grover,
Y.-C. He,
M. Metlitski,
R. Myers,
A. Sandvik,
T. Senthil,
A. Vishwanath,
W. Witczak-Krempa,
and
C. Xu.
This research was made possible by the
Natural Sciences and Engineering Research Council of Canada (NSERC) and by the facilities of the Shared Hierarchical 
Academic Research Computing Network (SHARCNET:www.sharcnet.ca) and Compute/Calcul Canada.
Research at Perimeter Institute is supported in part by the Government of Canada through the Department of Innovation, Science and Economic Development Canada and by the Province of Ontario through the Ministry of Economic Development, Job Creation and Trade.

\bibliographystyle{apsrev4-1}
\bibliography{bibliography.bib}

\end{document}